\documentclass[12pt]{article}

\usepackage{amstex}
\input epsf

\numberwithin{equation}{section}

\newcommand{\BH}{\boldsymbol{H}}
\newcommand{\BA}{\boldsymbol{A}}
\newcommand{\Ba}{\boldsymbol{a}}
\newcommand{\Bg}{\boldsymbol{g}}
\newcommand{\Bq}{\boldsymbol{q}}
\newcommand{\Bzero}{\boldsymbol{0}}
\newcommand{\Balpha}{\boldsymbol{\alpha}}
\newcommand{\Bbeta}{\boldsymbol{\beta}}
\newcommand{\Bvphi}{\boldsymbol{\varphi}}
\newcommand{\Btheta}{\boldsymbol{\theta}}
\newcommand{\Bomega}{{\boldsymbol{\omega}}}

\newcommand{\noi}{\noindent}

\begin{document}

\addtolength{\baselineskip}{2pt}
\thispagestyle{empty}

\begin{flushright}
SWAT/165\\
{\tt hep-th/9710073}\\
October 1997
\end{flushright}

\vspace{2.5cm}

\begin{center}
{\scshape\Large Strong Coupling $N=2$ Gauge Theory\\
\vspace{0.2cm}
with Arbitrary Gauge Group}

\vspace{1.5cm}

{\scshape\large Timothy J. Hollowood}

\vspace{0.5cm}
{\sl Department of Physics, University of Wales Swansea,\\
Swansea, SA2 8PP, U.K.}\\
{\tt t.hollowood@@swansea.ac.uk}\\

\vspace{1cm}

{\Large ABSTRACT}

\vspace{0.3cm}

\end{center}

A explicit definition of the cycles, on the auxiliary Riemann surface defined
by Martinec and Warner for describing pure $N=2$ gauge theories with 
arbitrary group, is provided. The strong coupling monodromies around
the vanishing cycles are shown to arise from a set of dyons which
becomes massless at the singularities. 
It is shown how the correct weak coupling monodromies are
reproduced and how the dyons have charges which are consistent with
the spectrum that can be calculated at weak coupling using
conventional semi-classical methods. 
In particular, the magnetic charges are co-root
vectors as required by the Dirac-Schwinger-Zwanziger quantization condition. 

\newpage

\section{Introduction}

A huge leap forward in the understanding of $N=2$ gauge theories in
the coulomb phase was initiated by Seiberg and Witten \cite{SW}. Their original
papers dealt only with the case of gauge group SU(2). Since then there
have been a number of papers setting out the generalization to
arbitrary gauge group. In particular, Martinec and Warner \cite{MW} 
develop a general construction of the auxiliary Riemann surface
which arise in Seiberg and Witten's approach. (Their work subsumes
earlier piece-meal generalizations to certain gauge groups, 
\cite{KLY,AF,KLT,BL,DS1,AS} 
although, it is not equivalent to the hyper-elliptic approach put forward
in \cite{DS2,AAG} for some groups, an approach that has been shown to be
incorrect for other reasons \cite{LPG}.) In the Martinec-Warner 
approach, the Riemann
surface is the spectral curve of an integrable system---the Toda
equations based on an associated affine algebra. In principle, this
relation to the integrable system allows one to extract the quantities
in the low energy effective action of the $N=2$ supersymmetric gauge
theory. More specifically, the cycles on the surfaces that are needed
to provide the solution of the low energy effective action are
determined by a special Prym subvariety of the Jacobian of the
surface. One goal of this paper is to provide an explicit construction
of these preferred cycles for any gauge group. The primary motivation is to use
these explicit expressions to subject the construction for arbitrary
gauge groups to the same rigorous tests as the
SU(2) case. In particular, we shall show how the strong coupling 
monodromies arise from renormalization around massless dyons 
singularities and how the
strong coupling monodromies reproduce all the
weak monodromies that can be calculated in perturbation
theory. As a by-product, we shall derive the electric and magnetic
charges of the
dyons that drive the strong coupling dynamics. This allows for another
highly non-trivial test of the construction because we can show that
these dyons are indeed in the spectrum of the theory at weak coupling
where conventional semi-classical methods are available for computing
the dyon spectrum. The result of this work is that the physics of the 
pure $N=2$ gauge theories with arbitrary gauge groups is now placed 
on a much firmer footing. 

Seiberg and Witten's approach determines the
exact prepotential of the low energy effective action of gauge theories
with $N=2$ supersymmetry. BPS states carry charges $Q=(\Bg,\Bq)$ 
and have a mass which is determined by the prepotential
\begin{equation}
M_Q=\left\vert Z_Q\right\vert=\left\vert Q\cdot A\right\vert,\label{eq:E1}
\end{equation}
where $A=(\Ba,\Ba_D)$ is a function on the
moduli space of vacua ${\mathcal M}$ and
the symplectic inner product is defined as
\begin{equation}
Q\cdot A=Q\Omega A^T=\Ba_D\cdot\Bg-\Ba\cdot\Bq\qquad
\Omega=\begin{pmatrix} 0&1\\ -1&0\end{pmatrix}\label{eq:E2}
\end{equation}
The vector $\Bg$ is the magnetic charge of the BPS state with respect
to the unbroken ${\rm U}(1)^r$ gauge group. It is a topological charge and 
is consequently quantized such that, with a suitable choice of overall
normalization, it is a vector of the co-root lattice $\Lambda_R^\vee$
of $g$, the (complexified) Lie algebra of the gauge group $G$. 
This is the lattice spanned by the simple co-roots
$\Balpha_i^\vee$, where we define the dual operation
\begin{equation}
\Balpha^\vee=2\Balpha/\Balpha^2.\label{eq:E4}
\end{equation}
The vector $\Bq$ determines the electric charge of the BPS state. It
not quite the true electric charge of the state, with respect to
the unbroken ${\rm U}(1)^r$ gauge group, because of theta-like terms
in the low energy effective action \cite{TJH4}. 
More precisely, it is Noether's charge
corresponding to global ${\rm U}(1)^r$ gauge transformations. 
Nevertheless, with a slight abuse of
language, we shall refer to $\Bq$ as the electric charge. This charge
is also quantized since the abelian group is embedded in the gauge
group $G$. This means that the allowed electric charge
vectors $\Bq$ must lie in the weight lattice of $g$. However, in a
pure gauge model, where all the fields are adjoint-valued,
only charges in the root lattice $\Lambda_R$ are actually realized.
To summarize
\begin{equation}
(\Bg,\Bq)\in\left(\Lambda_R^\vee,\Lambda_R\right).\label{eq:E5}
\end{equation}
Notice that the charges satisfy the generalized
Dirac-Schwinger-Zwanziger quantization condition:
\begin{equation}
Q_1\cdot Q_2=Q_1\Omega
Q_2^T=\Bg_1\cdot\Bq_2-\Bg_2\cdot\Bq_1\in{\mathbb Z}.
\end{equation}

Conventional electro-magnetic duality transformations
act on the fields in the low energy effective action. These
transformations induce an action on $A$ of the form
$A\rightarrow AD$, where $D$ is some matrix acting to the left. The
transformation takes the low energy effective action to an identical
action in the dual variables provided that $D$ is a symplectic 
transformation, i.e.
\begin{equation}
D\Omega D^T=\Omega.\label{eq:E6}
\end{equation}
It is easy to see that the BPS mass formula is invariant under this
transformation if the charges transform as $Q\rightarrow QD$.

The goal of the analysis is determine $A$. This is a multi-valued
function on ${\mathcal
M}$ since there are non-trivial monodromies along paths which encircle
certain co-dimension two
subspaces. Physically, the monodromy is causes by a certain BPS state
becomes massless on the subspace, causing a
logarithmic running of the effective coupling, in a way that we
calculate in section 2. A monodromy transformation around a cycle $C$
acts as $A\rightarrow AM$ and they are 
duality transformations since $M^T\Omega M=\Omega$.
This implies that a BPS state $Q$ taken around the cycle will end up as
the state $QM^{-1}$, unless the state passes across a surface on which
it decays to other BPS states of the same total charge.

Seiberg and Witten's major insight was that the multi-valued
function $A$ is determined solely in terms of a conjectured set of
the co-dimension two singular subspaces on which certain BPS states
become massless along with their associated monodromies. 
For SU(2) they surmised that there were two such
singularities, but in general in turns out that there are
$2r$ singularities corresponding to a set of $2r$ BPS states
$Q_i^a$, $i=1,\ldots,r$ and $a=1,2$. Although, this is a only a subset
of all the BPS states there is a `democracy of dyons', as first
described in the SU(2) case, whereby any dyon in the (semi-classical) 
spectrum becomes
massless at one of the singularities by following a path to the
singularity with some appropriate non-trivial monodromy which
transforms it to one of the set $Q_i^a$.
The construction proceeds by defining
a Riemann surface with a moduli space identified with
${\mathcal M}$. For the surface,
there exists a special meromorphic 
one-form $\lambda$ and a set of $2r$ preferred
homology one-cycles $\nu_i^a$, such that
\begin{equation}
Z_{Q_i^a}=Q_i^a\cdot A=\oint_{\nu_i^a}\lambda.\label{eq:E7}
\end{equation}
Since the charges $Q_i^a$ are linearly independent these relations are enough
to determine the function $A$ and a mapping between charges $Q$ and
cycles $C_Q$. The picture is now clear: a subspace
where a dyon $Q$ becomes massless is precisely the subspace on which
the cycle $C_Q$ vanishes and the Riemann surface degenerates.

Rather than begin with a conjectured set of charges for the massless
BPS states at the singularities and then proceed to construct the
Riemann surface and the other data, as Seiberg and Witten did in the
SU(2) case, we shall start with the Riemann
surface data and show that resulting function $A$ has the correct
monodromy properties which agree with calculations performed in
dual perturbation theory in the neighbourhood of the singularities and at
weak coupling in conventional perturbation theory.

\section{The Monodromies in Perturbation Theory}

In this section, we calculate the monodromies of the function
$A$. At weak coupling these may be calculated using standard
perturbation theory. At strong coupling, it is conjectured that the
monodromies arise from paths around singularities on which certain
dyons become massless, and consequently these monodromies can also be 
calculated in perturbation theory, but now in dual variables.

At weak coupling the moduli space of vacua can be parameterized by the
classical Higgs VEV $\Phi$, modulo global gauge transformations. These
transformations can always be used to conjugate $\Phi$ into the Cartan
subalgebra $\Ba\cdot\BH$, which defines the complex $r$-dimensional
vector $\Ba$. The remaining freedom to perform conjugations in
the Weyl group of $g$, can be used to choose, say, ${\rm Re}(\Ba)$ to
be in the fundamental Weyl chamber of $g$, i.e.
\begin{equation}
\Balpha_i\cdot{\rm Re}(\Ba)\geq0,\label{eq:E8}
\end{equation}
where $\Balpha_i$, $i=1,\ldots,r$, are the set of simple roots of $g$.
On the wall of the fundamental Weyl chamber, where $\Ba\cdot\Balpha_i=0$,
points are identified under $\sigma_i$, the Weyl reflection in $\Balpha_i$.
At weak coupling we parameterize ${\mathcal M}$ by $\Ba$ so
constrained.

The function $\Ba_D$ can be calculated as a function of the Higgs VEV,
i.e.~$\Ba$, in perturbation theory. The calculation is standard and to
one-loop 
\begin{equation}
\Ba_D={i\over2\pi}\sum_{\Bbeta}\Bbeta(\Bbeta\cdot\Ba)\ln\left({
\Bbeta\cdot\Ba\over\Lambda}\right),\label{eq:E9}
\end{equation}
where $\Lambda$ is the usual strong coupling scale and the sum is over all
the root of $g$. The perturbative regime
is valid as long as $|\Ba\cdot\Balpha_i|\gg\Lambda$, for
$i=1,\ldots,r$. It is immediately apparent that $\Ba_D$ is not
single-valued as one follows a cycle around one of the $r$ co-dimension two
subspaces $\Ba\cdot\Balpha_i=0$. Bearing in mind that such a cycle
involves an identification by the Weyl reflection in the simple root
$\Balpha_i$,  on the
wall ${\rm Re}(\Ba)\cdot\Balpha_i=0$, one finds that the monodromy,
for some choice of orientation, is $A\rightarrow AM_i$, where \cite{KLT,DS1,BL}
\begin{equation}
M_i=\begin{pmatrix} \sigma_i&\Balpha_i\otimes\Balpha_i\\ 0&\sigma_i
\end{pmatrix}.\label{eq:E10}
\end{equation}
Notice that $M\Omega M^T=\Omega$, so that the monodromy transformation
is a duality transformation. It is important that non-perturbative
corrections to $\Ba_D$ do not make any additional contributions to 
weak coupling monodromies. 
The monodromy transformations generate a representation of the Brieskorn
Braid Group.

It will turn out that the
weak coupling monodromies $M_i$ are the remnant of a larger set of
monodromies that are present at strong coupling. These monodromies
arise from following cycles around co-dimension two subspace on which
a certain BPS state $Q$ becomes massless, i.e. $Z_Q=0$. The
resulting monodromy $M_Q$ around the singularity can be calculated by
performing a duality transformation to dual variables, which are local
with respect to the BPS state, and then using perturbation theory in those dual
variables. The result was written down in \cite{KLT}:
\begin{equation}
M_Q=1+\left(\Omega Q^T\right)Q.\label{eq:E11}
\end{equation}
Notice that $M_Q\Omega M_Q^{-1}=\Omega$ so that the monodromy
transformation is a duality transformation.
For completeness we will sketch the proof of this relation.

For any given charge $Q$, there always exists some duality
transformation $D$ such that
\begin{equation}
Q'=QD=\left(\Bzero,\Bq'\right),\qquad A'=AD.\label{eq:E12}
\end{equation}
So in the transformed frame the state is purely electrically
charged. The effective field theory in the vicinity of the subspace
on which the state becomes massless consists of a vector
supermultiplet containing a set of photons
which are related by the duality transformation $D$ to the photons of
the unbroken ${\rm U}(1)^r$ symmetry, and a light hypermultiplet of 
electric charge $\Bq'$ describing the BPS state. 
These light charged states cause the dual coupling
constant to run in an asymptotically infra-red free way. To one-loop
perturbation theory in the dual variables one has 
\begin{equation}
\Ba'_D=-{i\over2\pi}\Bq'\left(\Bq'\cdot\Ba'\right)\ln
\left({\Bq'\cdot\Ba'\over\Lambda'}\right).\label{eq:E13}
\end{equation}
This one-loop expression is enough to determine the monodromy around
the singularity $Z_Q=\Ba'\cdot\Bq'=0$:
\begin{equation}
A'\rightarrow A'M_{Q'},\qquad M_{Q'}=\begin{pmatrix}
1&\Bq'\otimes\Bq'\\ 0&1\end{pmatrix}=1+\left(\Omega Q^{\prime
T}\right)Q'.\label{eq:E14}
\end{equation}

To find the monodromy transformation $M_Q$ we simply have to transform
back to the original variables $M_Q=D^{-1}M_{Q'}D$, which given that
$D$ is a duality transformation, yields
the expression in \eqref{eq:E11}. Using \eqref{eq:E11}, 
the action of the monodromy
transformation $M_Q$ on the charge $\tilde Q$ is
\begin{equation}
\tilde Q\rightarrow \tilde QM_Q=\tilde Q+(\tilde Q\cdot Q)Q.\label{eq:E15}
\end{equation}

\section{The Riemann Surface: Simply-Laced}

In this section, we explain how the Riemann surface 
and the associated data $\lambda$ 
and the $2r$ preferred cycles  $\nu_i^a$ are constructed.
It is helpful to consider the simply-laced and non-simply-laced cases
separately and so in this section we shall be considering the 
former cases only, the non-simply-laced cases will considered in
a later section. 

When the gauge group $G$ has a simply-laced Lie algebra
$g$ one chooses a representation $\rho$
of $g$ and defines a Riemann surface   
via a characteristic polynomial in two auxiliary variables $x$ and $z$:
\begin{equation}
{\rm det}\left[\rho(\boldsymbol{A}(z))-x\cdot\boldsymbol{1}\right]=0,
\label{eq:E16}
\end{equation}
where
\begin{equation}
\boldsymbol{A}(z)=\Bvphi\cdot\BH+\sum_{i=1}^r\left(E_{\Balpha_i}+
E_{-\Balpha_i}\right)+zE_{-\Btheta}+\mu z^{-1}E_{\Btheta}.
\label{eq:E17}
\end{equation}
In the above, $E_{\Balpha_i}$ is the step generator of $g$ associated
to the simple root $\Balpha_i$ and $E_{\Btheta}$ is the generator
associated to the highest root $\Btheta$. The Cartan
generators of $g$ are denoted by the $r={\rm rank}(g)$ dimensional
vector $\BH$. The $r$-dimensional complex vector $\Bvphi$ parametrizes the 
moduli space of the surface which is identified with $\cal M$. 
The parameter $\mu$
is equal to $\Lambda^{2{\rm dim}(\rho)}$, where 
$\Lambda$ is the familiar scale of strong coupling effects. It turns
out that the surface that has been defined is the spectral curve of an
integral system. In fact 
$\BA(z)$ is the Lax operator of the $g^{(1)}$
Toda system, with $z$ playing the role of the loop variable \cite{MW}.
This fascinating observation will not play any central role in the
present discussion.

Following \cite{MW}, it is most convenient to think of the 
Riemann surface as a foliation over the Riemann sphere for
$z$ by extending $x$ to an analytic function of $z$. The number of
leaves of the foliation is
then equal to the dimension of the representation $\rho$. The operator
$\BA(z)$ is always conjugate to some element of the Cartan subalgebra
$\tilde{\Bvphi}(z)\cdot\BH$, however, there remains the freedom to
to perform conjugations within the Weyl subgroup of the gauge group.
This freedom can be fixed by choosing, for example, ${\rm
Re}(\tilde{\Bvphi}(z))$ to
be in, or on the wall of, the Fundamental Weyl Chamber, i.e.
\begin{equation}
\Balpha_i\cdot{\rm Re}(\tilde{\Bvphi}(z))\geq0.
\label{eq:E171}
\end{equation}
This condition determines a series of cuts on the $z$-plane.

On the leaf of the foliation associated to the weight vector
$\Bomega$, we have
$x=\Bomega\cdot\tilde{\Bvphi}(z)$. Two leaves are connected at
branch-points whenever two such
eigenvalues of $\BA(z)$ coincide. Notice that due to our choice
\eqref{eq:E171}, this can only happen when 
\begin{equation}
\Balpha_i\cdot\tilde{\Bvphi}(z)=0,
\label{eq:E172}
\end{equation}
for a simple root
$\Balpha_i$. At such a point,
the pair of sheets labelled by $\Bomega$ and $\Bomega'$, for which
$\Bomega'=\sigma_i\left(\Bomega\right)$,
are joined. (Here, $\sigma_i$ is the Weyl reflection in the
simple root $\Balpha_i$.) It follows from this, that the foliation
splits into disconnected components corresponding to the separate orbits
of the weights under the Weyl group of $g$. 
Since $\rho\left(\boldsymbol{A}(z)\right)^T=\rho
\left(\boldsymbol{A}(\mu/z)\right)$, it
follows that \eqref{eq:E16} is invariant under $z\rightarrow\mu/z$.
As a consequence, the branch-points will occur in pairs $z_i^\pm$, related
by $z_i^+z_i^-=\mu$. To fix the definition, we define 
$z_i^-$ to be the branch-points
that tend to $0$ in the weak coupling limit $\mu\rightarrow0$.
There are also branch-points at $0$ and $\infty$ that 
connect all the sheets corresponding to 
weights on a particular orbit of the Weyl group.
The branch-points are connected by cuts which are specified by our
Weyl Group-fixing choice in \eqref{eq:E171}.  

Notice that
the positions of the branch-points are independent of the
representation $\rho$. Clearly this will prove significant when we
come to show that the construction is independent of the choice of
representation $\rho$. Notice also, that the lift of the 
contour, labelled $C$ in Fig.~1, which encircles the origin
and the branch-points $z_i^-$, will return to the same sheet after one
circuit.

\begin{figure}
\begin{center}
\leavevmode \epsfxsize=10cm \epsfbox{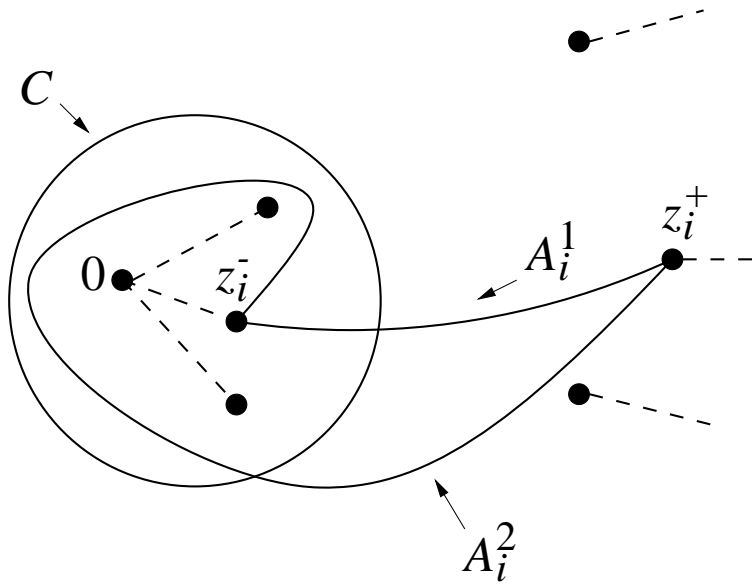}
\end{center}
\caption{Definition of contours $A_i^a$ and $C$}
\end{figure}

As alluded to in the introduction, the connection between the 
Riemann surface
and the $N=2$ supersymmetric gauge theories is established through the
existence of a special meromorphic 
differential $\lambda$ and a preferred set of
$2r$ cycles on the surfaces. The differential is simply
\begin{equation}
\lambda={1\over2\pi i}{xdz\over z},\label{eq:E19}
\end{equation}
which, due to dependence on $x$, has a different expression on each sheet.

The vanishing cycles correspond to points in the moduli space where a
pair of branch-points $z_i^\pm$ become coincident. Obviously since
$z_i^-z_i^+=\mu$, this occurs at either $z=\pm\sqrt\mu$. A given pair of
branch-points can come together along an infinite number paths
depending on the arbitrary number of circuits around the origin that
are made before the points become coincident. In order to specify the
solution, it is necessary to choose just two vanishing paths, for each
$i$, one associated to when
$z_i^\pm$ come together at $z=\sqrt\mu$, and the other at $-\sqrt\mu$. 
This will then provide the $2r$ cycles required for defining $\Ba$ and
$\Ba_D$. The $2r$ contours will be denoted by 
$A_i^a$, $i=1,\ldots,r$ and $a=1,2$. For fixed $i$, each pair
joins $z_i^-$ to $z_i^+$ along a path which avoids any cut on the same
sheet of the foliation as $z_i^\pm$. 
There is a certain ambiguity present in the
choice of the contours which we will later interpret as a manifestation
of the `democracy of dyons' described in the context of SU(2) in \cite{SW}.
Let us suppose we choose a set of contours
$A_i^1$ with intersections defined by the anti-symmetric matrix ${\cal
I}_{ij}$. We will fix our definition of orientation by saying that
the intersection of two contours $A$ and $B$, denoted $A\circ B$ is $1$
($-1$) if $B$ crosses $A$ from right (left).
It is always possible to choose, for example, ${\cal
I}_{ij}=0$ when ever $\Balpha_i\cdot\Balpha_j\neq0$, which fixes the
ambiguity up to an
overall integer corresponding to the number of times the contours
as a whole wind around the origin. The partner contours
$A_i^2$ are then defined as 
\begin{equation}
A_i^2=C+A_i^1,
\end{equation}
i.e. with an extra winding around the origin. This is illustrated in
Fig.~1. It is easy to see that
the pair $A_i^a$, for fixed $i$, correspond to paths where $z_i^\pm$
come together at each of the two singular points $\pm\sqrt\mu$. 
For example, Fig.~2 shows
the branch-points and a choice of contours $A_i^1$ for $g=A_3$. In
this case ${\cal I}_{12}={\cal I}_{13}={\cal I}_{23}=1$.

\begin{figure}
\begin{center}
\leavevmode \epsfxsize=8cm \epsfbox{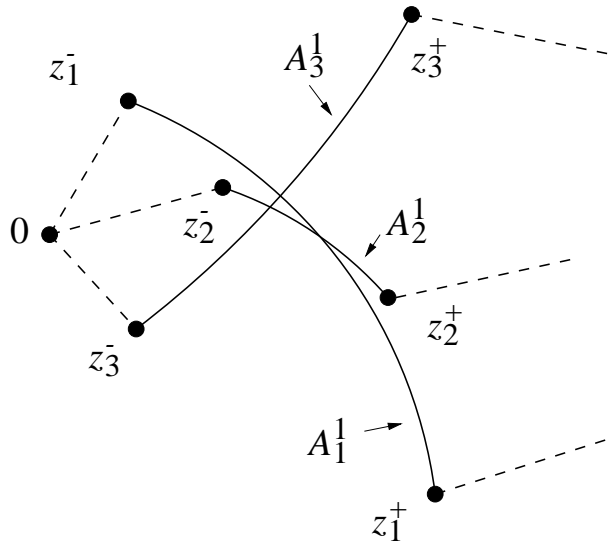}
\end{center}
\caption{Choice of contours $A_i^1$ for $g=A_3$}
\end{figure}

The set of $2r$ cycles $\nu_i^a$ that are used to specify the
solution in \eqref{eq:E7} 
are defined as particular lifts of the contours $A_i^a$ to the
leaves of the foliation. The weighting given to each contour will turn
out to be crucial and we shall find that the correct lift is
\begin{equation}
\nu_i^a={1\over N_\rho}\sum_{\Bomega}
\left(\Bomega\cdot\Balpha_i\right)A_i^a(\Bomega),\label{eq:E20}
\end{equation}
where $A_i^a(\Bomega)$ is the lift of the contour $A_i^a$ to the
sheet labelled by $\Bomega$ and $N_\rho$ is a normalization factor
which will be determined below. Before we continue it is important to
show that the $\nu_i^a$ are indeed (closed) one-cycles on the
surface. The proof becomes trivial when one notices that contour
$A_i^a(\Bomega)$ joining $z_i^-$ to $z_i^+$ on the sheet
associated to $\Bomega$ is always accompanied by a return
contour $A_i^a(\Bomega')$, having opposite weight in
\eqref{eq:E20}, on the sheet
associated to $\Bomega'=\sigma_i\left(\Bomega\right)$.
Hence the $\nu_i^a$ are indeed closed one-cycles. In the following,
it is convenient to make this explicit by introducing
the closed cycles
\begin{equation}
\hat A_i^a(\Bomega)=A_i^a(\Bomega)-A_i^a\left(\sigma_i(\Bomega)\right),
\end{equation}
in terms of which
\begin{equation}
\nu_i^a={1\over 2N_\rho}
\sum_{\Bomega}\left(\Bomega\cdot\Balpha_i\right)
\hat A_i^a(\Bomega).
\end{equation}
The extra factor of $1/2$ is to prevent over-counting due to the fact that 
$\hat A_i(\sigma_i(\Bomega))=-\hat A_i(\Bomega)$.
It is now a simple matter to show that $\Ba$ and $\Ba_D$ are 
independent of the representation
$\rho$, once the normalization factor $N_\rho$ has been fixed. We
first notice that
\begin{equation}
\oint_{\hat A_i^a(\Bomega)}\lambda=\int_{A_i^a}
\left(\Bomega\cdot\tilde{\Bvphi}(z)-
\Bomega'\cdot\tilde{\Bvphi}(z)\right){dz\over z}
={\Bomega\cdot\Balpha_i\over\Balpha_i^2}\int_{A_i^a}\Balpha_i
\cdot\tilde{\Bvphi}(z){dz\over z}\label{eq:E213}
\end{equation}
This means that $\oint_{\nu_i^a}\lambda$ only depends on the
representation $\rho$ through the overall normalization
$\sum_{\Bomega}(\Bomega\cdot\Balpha_i)^2/N_\rho$. By taking
\begin{equation}
N_\rho\cdot{\mathbf 1}=\sum_{\Bomega}\Bomega\otimes\Bomega,\label{eq:E29}
\end{equation}
guarantees that the integrals of $\lambda$ around the $\nu_i^a$ and
hence $\Ba$ and $\Ba_D$ are independent
of $\rho$. The cycles that we have constructed pick out the special
Prym subvariety of the Jacobian of the surface. It is precisely the
subvariety associated to the reflection representation of the Weyl
group \cite{MW}. This is explicit in \eqref{eq:E20}, where the Weyl
group acts on the contours as $\sigma:\ A_i(\Bomega)\mapsto
A_i(\sigma(\Bomega))$.

In order to make connection with the electric and magnetic charges, it
is useful to notice that in the weak coupling limit, although the
integrals of $\lambda$ around $\nu_i^a$ are logarithmically
divergent, the integral around the difference
$\nu_i^1-\nu_i^2$, which is a lift of the cycle $C$, 
is well-defined. To see this, notice that
in the weak coupling limit $\mu\rightarrow0$ so
all the branch-points $z_i^-\rightarrow0$.
At $\mu=0$, the function $\tilde{\Bvphi}(z)$ is analytic at
the origin and therefore using \eqref{eq:E213} we find that the
integral is equal to
$\Balpha_i\cdot\tilde{\Bvphi}_{\mu=0}(0)$.
We can therefore identify $\tilde{\Bvphi}_{\mu=0}(0)$ as the vector
which specifies the classical Higgs VEV, that is $\Ba$ in the weak 
coupling limit. Hence $\nu_i^1-\nu_i^2$ is a purely electric cycle
corresponding to an electric charge vector $\Balpha_i$. This means
that the charges of the dyons satisfy:
\begin{equation}
Q_i^1-Q_i^2=(\Bzero,\Balpha_i).\label{eq:E101}
\end{equation}

\section{The Strong Coupling Monodromies: Simply-Laced}

In this section, we will show that the monodromy transformations around
the $2r$ vanishing cycles $\nu_i^a$ 
can be identified with the monodromies associated to a set of $2r$ BPS
states, as calculated in section 2. Once we have identified these 
states we can then calculate the weak coupling monodromies and show that they
reproduce those calculated in section 2.

In order to calculate the monodromy transformations around a vanishing
cycle, we use the Picard-Lefshetz Theorem which allows one to calculate the
action of monodromies on the cycles $\nu_i^a$ 
themselves. We can then lift this
action to the charges $Q_i^a$ by using \eqref{eq:E7}.
The Picard-Lefshetz Theorem states that the monodromy
associated to a vanishing cycle with a single connected component
$\delta$ on another cycle is
\begin{equation}
{\mathcal M}_\delta(\zeta)=\zeta+(\zeta\circ\delta)\delta.\label{eq:E31}
\end{equation}
Here, $\zeta\circ\delta$ is the intersection number of the two cycles.
Since our vanishing cycles have, in general, more than one connected
component we will need a refinement of the Picard-Lefshetz Theorem. If
$\delta=\sum_i n_i\delta_i$, a union of disconnected components with
certain non-zero coefficients $n_i$, then 
\begin{equation}
{\mathcal M}_\delta(\zeta)=\zeta+\sum_i(\zeta\circ\delta_i)\delta_i.
\label{eq:E32}
\end{equation}

Applying this formula allows us to deduce the action of the 
monodromies ${\mathcal M}_{\nu_i^a}$ associated to the vanishing cycles
$\nu_i^a$ on the other vanishing cycles. 
Before we do this, it is helpful to know
that 
\begin{equation}
\hat A_i^a(\Bomega)\circ
\hat A_j^b(\Bomega')=\left(\epsilon_{ab}+{\cal I}_{ij}\right)
\left(\delta_{\Bomega,\Bomega'}-\delta_{\sigma_i(\Bomega),
\Bomega'}-\delta_{\Bomega,\sigma_j(\Bomega')}+\delta_{\sigma_i(
\Bomega),\sigma_j(\Bomega')}\right).\label{eq:E33}
\end{equation}
In the above, ${\cal I}_{ij}$ is the
intersection form of the $A_i^1$ contours in the $z$-plane defined previously
and $\epsilon_{ab}$ is the
two-dimensional anti-symmetric tensor.

Using \eqref{eq:E32} and \eqref{eq:E33},
we can now calculate the monodromy transformations
around the vanishing cycles:
\begin{equation}
{\mathcal M}_{\nu_i^a}\left(\nu_j^b\right)=
\nu_j^b+{1\over2}\sum_{\Bomega}
\left(\nu_j^b\circ\hat A_i^a(\Bomega)\right)
\hat A_i^a(\Bomega).\label{eq:E35}
\end{equation}
The extra factor of $1/2$ is to prevent over-counting due to
$\hat A_i(\sigma_i(\Bomega))=-\hat A_i(\Bomega)$.
The intersection number above can be written as
\begin{align}
\nu_j^b\circ\hat A_i^a(\Bomega)
&={1\over 2N_\rho}\sum_{\Bomega'}
\Bomega'\cdot\Balpha_j\left(
\hat A_j^b(\Bomega')\circ\hat A_i^a(\Bomega)\right) \notag \\
&={1\over N_\rho}\left(\epsilon_{ba}+{\cal I}_{ji}
\right)\left(\Balpha_j\cdot\Bomega-
\Balpha_j\cdot\sigma_i(\Bomega)\right) \label{eq:E36} \\
&={1\over N_\rho}
\left(\epsilon_{ba}+{\cal I}_{ji}
\right){2\Balpha_i\cdot\Balpha_j\over\Balpha_i^2}
\Balpha_i\cdot\Bomega,\notag
\end{align}
where we have used \eqref{eq:E33}. Since $g$ is simply-laced, all the
roots have the same length which we normalize as $\Balpha_i^2=2$, in
which case \eqref{eq:E35} becomes
\begin{equation}
{\mathcal M}_{\nu_i^a}\left(\nu_j^b\right)=
\nu_j^b+\left(\epsilon_{ba}+{\cal I}_{ji}\right)
\left(\Balpha_i\cdot\Balpha_j\right)\nu_i^a.\label{eq:E361}
\end{equation}
This is equivalent, via \eqref{eq:E7}, to the following 
action on the charges $Q_i^a$:
\begin{equation}
{\mathcal M}_{Q_i^a}\left(Q_j^b\right)=
Q_j^b+\left(\epsilon_{ba}+{\cal I}_{ji}\right)\left(\Balpha_i\cdot
\Balpha_j\right)Q_i^a.\label{eq:E362}
\end{equation}

By comparing \eqref{eq:E362} with \eqref{eq:E15} we see that the charges must
have the following symplectic inner products:
\begin{equation}
Q_j^b\cdot
Q_i^a=\left(\epsilon_{ba}+{\cal I}_{ji}\right)\Balpha_i\cdot\Balpha_j.
\label{eq:E38}
\end{equation}
The charges are further constrained by \eqref{eq:E101}. 
This determines the charges in the form
\begin{equation}
Q_i^1=\left(\Balpha_i,p_i\Balpha_i\right),\qquad
Q_i^2=\left(\Balpha_i,(p_i+1)\Balpha_i\right),\label{eq:E39}
\end{equation}
where, for $i$ and $j$ such that $\Balpha_i\cdot\Balpha_j\neq0$,
\begin{equation}
p_i-p_j={\cal I}_{ji},
\end{equation}
otherwise $p_i-p_j$ is unconstrained. Since the Dynkin diagrams have no closed
loops, these equations always admit a
solution for integer $p_i$, up to an overall integer. 
The integer $p_i$ is correlated
with the choice of the contour $A_i^1$ in the sense that if we shift
it by an additional winding around the origin, i.e.~$A_i^1\rightarrow A_i^1+C$,
then $p_i\rightarrow p_i+1$. The overall integer reflects the freedom
for all the
contours to wind around the origin an arbitrary number of
times. Recall that it is always possible to choose the $A_i^1$
contours so that ${\cal I}_{ji}=0$ whenever
$\Balpha_i\cdot\Balpha_j\neq0$. With this choice we have $p_i=n$,
for some integer $n$, for all $i$. For the $A_3$ example in Fig.~2 the
charges are
\begin{equation}
Q_1^1=\left(\Balpha_1,n\Balpha_1\right),\qquad Q_2^1=
\left(\Balpha_2,(n+1)\Balpha_2
\right),\qquad Q_3^1=\left(\Balpha_3,(n+2)\Balpha_3\right),
\end{equation}
for integer $n$.

From the above, we deduce that the dyon of charge 
$(\Balpha_i,n\Balpha_i)$ has a vanishing cycle
that is the lift of $A_i^1+(n-p_i)C$. This is a manifestation of the
democracy of dyons.

\section{The Riemann Surface: Non-Simply-Laced}

In this section, we explain how to generalize the discussion to 
the theories whose gauge group has a non-simply-laced Lie algebra. 
Many of the details are similar to the simply-laced cases.

First of all, we define the dual algebra
$g^\vee$ to be the algebra obtained by interchanging long and short
roots, i.e.~by replacing each root $\Balpha$ by its co-root 
$\Balpha^\vee=2\Balpha/\Balpha^2$. The construction of the
characteristic polynomial for a non-simply-laced algebra $g$, involves
a simply-laced algebra $\tilde g$ related to $g$ by $\tilde
g^{(\tau)}=\left(g^{(1)}\right)^\vee$. In other words
the dual of the affine algebra $g^{(1)}$ is the
twisted affinization of the simply-laced algebra $\tilde g$ of
respect to an outer (diagram) automorphism $\pi$ with 
order $\tau$. For the non-simply-laced Lie algebras the associated
simply-laced algebras are given by
\begin{equation}
\left(B_r^{(1)}\right)^\vee=A_{2r-1}^{(2)},\quad
\left(C_r^{(1)}\right)^\vee=D_{r+1}^{(2)},\quad
\left(F_4^{(1)}\right)^\vee=E_6^{(2)},\quad
\left(G_2^{(1)}\right)^\vee=D_4^{(3)}.\label{eq:E21}
\end{equation}
Under the outer automorphism, $\tilde g$ is decomposed into eigenspaces:
\begin{equation}
\tilde g=\bigoplus_{p=0}^{\tau-1}\tilde g_p,\label{eq:211}
\end{equation}
where $\pi(\tilde g_p)=\exp(2\pi ip/\tau)\tilde g_p$.
The invariant subalgebra $\tilde g_0$ is precisely the original
non-simply-laced Lie algebra $g$ and the eigenspaces are irreducible
highest weight representations of $g$. The simple roots of $g$ are
identified with the $\pi$-invariant combinations of the simple-roots of
$\tilde g$ for each orbit of the simple-roots under $\pi$:
\begin{equation}
\Balpha_i=\sum_{p=1}^{\tau_i}\tilde{\Balpha}_{i_p},\label{eq:E22}
\end{equation}
where $\tilde{\Balpha}_{i_p}=\pi^{p-1}(\tilde{\Balpha}_{i_1})$, 
$p=1,\ldots,\tau_i$,
are the simple roots on the $i^{\rm th}$
orbit of the outer automorphism $\pi$. The orbit is either of
dimension $\tau$, in which
case $\Balpha_i$ is a long root of $g$, or one dimensional, in which
case $\Balpha_i$ is a short root of $g$. In the latter case, we have by
definition $i_p=i_1$. It is convenient to define $L$
and $S$ as the set of $i$ such that $\Balpha_i$ is long and short,
respectively. We will choose the labelling in
each of the long orbits so that
$\Balpha_{i_p}\cdot\Balpha_{j_q}\neq0$ only for $p=q$. 

The Riemann surface for $g$ is constructed as in
\eqref{eq:E16} by replacing $g^{(1)}$ by 
the twisted affine algebra $\tilde g^{(\tau)}$.  One fixes a representation
$\tilde\rho$ of $\tilde g$ and the $\tilde r={\rm rank}(\tilde g)$
dimensional complex vector $\Bvphi$ is constrained to lie in the
$r$-dimensional subspace invariant under the outer
automorphism $\pi$.\footnote{We will use the single symbol $\pi$ to describe
the various actions of the outer
automorphism: (i) on the simple roots (ii) lifted into the Lie algebra
(iii) on vectors in Cartan space, since no confusions should arise.}
The generator $E_{\Btheta}$ is now 
associated to $\Btheta$ the highest weight of the representation of 
$\tilde g_1$ of $g$, rather than the highest root. 
As in section 3, we shall think of the
Riemann surface as a foliation over the $z$ plane and we shall fix the details
of the foliation by choosing
\begin{equation}
\tilde{\Balpha}_{i_p}\cdot{\rm Re}\left(\tilde{\Bvphi}(z)\right)\geq0.
\end{equation}

There are branch-points connecting two sheets $\tilde{\Bomega}$ and
$\tilde{\Bomega}'=\sigma_{i_p}(\tilde{\Bomega})$ 
of the foliation when $\tilde{\Balpha}_{i_p}\cdot\tilde{\Bvphi}(z)=0$. 
These branch-points come in $\tau r$ pairs $z_i^\pm(p)$,
$i=1,\ldots,r$ and $p=1,\ldots,\tau$. To see this, we remark that
because $\tilde{\Balpha}_{i_p}\cdot\tilde{\Bvphi}(z)$ is an element of
$\tilde g^{(\tau)}$ it is invariant under a
combination of the action of the outer automorphism $\pi$ along with 
$z\rightarrow\exp(-2\pi i/\tau)z$, i.e.
\begin{equation}
\pi\left(\tilde{\Balpha}_{i_p}\right)\cdot\tilde{\Bvphi}
\left(z\exp(-2\pi i/\tau)\right)=\tilde{\Balpha}_{i_p}\cdot\tilde{\Bvphi}(z).
\end{equation}
Therefore the outer automorphism has an
action on the branch-points in the $z$-plane given by 
\begin{equation}
z_i^\pm(p+1)=\exp(2\pi i/\tau)z_i^\pm(p).\label{eq:E221}
\end{equation}
Notice that the branch-points associated to a simple root
$\tilde{\Balpha}_{i_1}$ which is fixed under $\pi$ come with a
multiplicity of $\tau$.
Since $\tilde\rho\left(\boldsymbol{A}(z)\right)^T=\tilde\rho
\left(\boldsymbol{A}'(\mu/z)\right)$, where the prime indicates a
re-labelling of the simple roots $i_p\leftrightarrow i_{-p}$ (where
the label $p$ is understood to be defined modulo $\tau$)
one easily deduces the additional relation
\begin{equation}
z_i^+(p)z_i^-(-p)=\mu.
\end{equation}

The preferred set of $2r$ vanishing
cycles $\nu_i^a$, $i=1\ldots,r$ and $a=1,2$  
are defined as follows. 
Each of the pairs $z_i^\pm(p)$ can come together at two points
$\pm\sqrt\mu\exp(2\pi ip/\tau)$ and for each pair we define two
contours in the $z$-plane 
$A_i^a(p)$, $a=1,2$, which vanish when $z^\pm_i(p)$ come
together at each of two points above. 
As for the simply-laced cases, there is a certain ambiguity
in defining contours. Firstly, we choose the set of contours 
such that, under $z\rightarrow z\exp(2\pi i/\tau)$, $A_i^1(p+1)$ is the image
of $A_i^1(p)$. Let us denote the intersection number of these
contours by ${\cal I}_{ij}(p-q\,{\rm mod}\tau)$. 
The fact that the intersection
form depends on $p-q$ alone is due to its invariance under the outer
automorphism. Each contour has a partner $A_i^2(p)$
defined in the following way. If the simple root
$\tilde{\Balpha}_{i_p}$ is not fixed under $\pi$
then 
\begin{equation}
i\in L:\qquad A_i^2(p)=C+A_i^1(p),
\end{equation}
where $C$ is the contour that surrounds the origin and all the
$z_i^-(p)$, i.e.~$A_i^2(p)$ is equal to $A_i^1(p)$ plus an extra
winding around the origin. If the
simple root $\tilde{\Balpha}_{i_p}$ is fixed under $\pi$, then 
$A_i^2(p)$ is the contour which connects $z_i^-(p)$ to
$z_i^+(p+1)$ (with $p$ defined modulo $\tau$) in
such a way that 
\begin{equation}
i\in S:\qquad A_i^2(p)=C(p)+A_i^1(p),
\end{equation}
where $C(p)$ is the contour which connects $z_i^+(p)$ to $z_i^+(p+1)$.
These contours lift to the foliation in an identical way to those in
section 3. We define $A_i^a(\tilde{\Bomega};p)$ to be the lift of $A_i^a(p)$
to the sheet labelled by the weight $\tilde{\Bomega}$. Notice the essential
difference between $A_i^2(\tilde{\Bomega};p)$, according to whether $i\in L$ or
$\in S$. In the latter case, the $z_+(p)$ are on the same sheets of the
foliation, hence it is consistent for a contour to emerge from
$z_i^-(p)$ and disappear down $z_i^+(p+1)$. This is not so when $i\in L$.

As an example consider the case of $G_2$ which arises via a third order
outer automorphism of $D_4$. There are two orbits for the simple roots
under $\pi$, $i=1$ containing $\tilde{\Balpha}_1$, $\tilde{\Balpha}_3$ and
$\tilde{\Balpha}_4$ and $i=2$ containing $\tilde{\Balpha}_2$
only. This is illustrated in Fig.~3.

\begin{figure}
\begin{center}
\leavevmode \epsfxsize=6cm \epsfbox{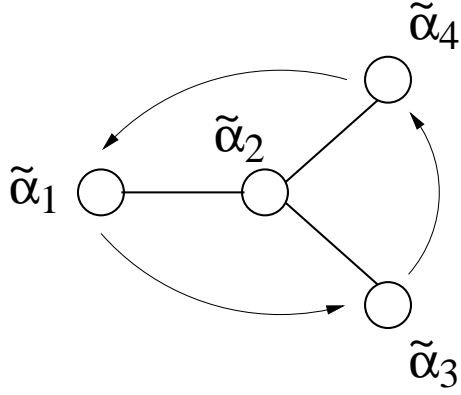}
\end{center}
\caption{Action of outer automorphism on $D_4$ giving $G_2$}
\end{figure}

\noi
The positions of the branch-points and a choice for the contours
$A_i^1(p)$ (along with only $A_2^2(3)$ for clarity) is illustrated in Fig.~4.
With this choice ${\cal I}_{12}(0)=-1$ and ${\cal I}_{12}(1)={\cal
I}_{12}(2)=0$. 

\begin{figure}
\begin{center}
\leavevmode \epsfxsize=10cm \epsfbox{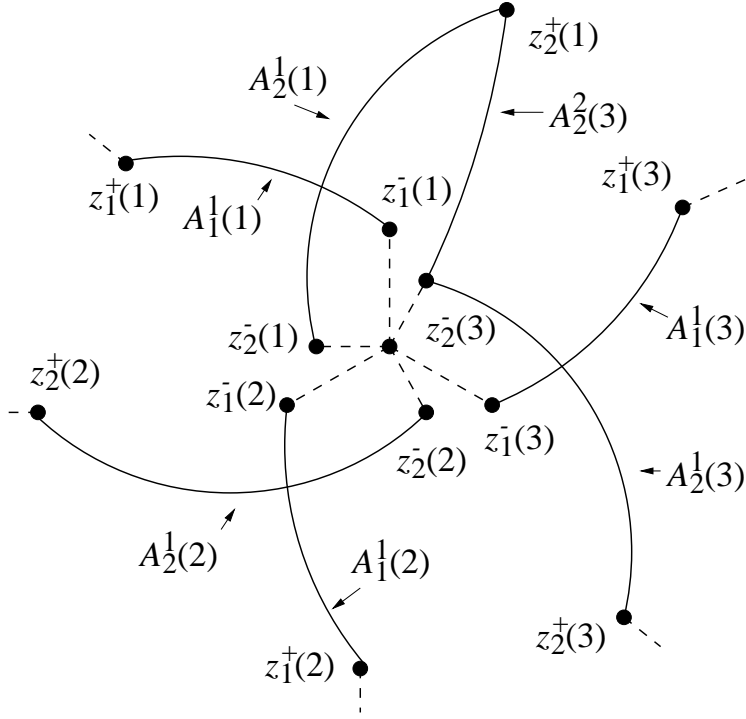}
\end{center}
\caption{Choice of contours for $G_2$}
\end{figure}

In terms of these cycles the basis set of 
preferred cycles are 
\begin{equation}
\nu_i^a={1\over N_{\tilde\rho}}\sum_{\tilde{\Bomega}}\sum_{p=1}^\tau
\left(\tilde{\Bomega}\cdot\pi^{p-1}(\tilde{\Balpha}_{i_1})\right)A_i^a
(\tilde{\Bomega};p).
\end{equation}
where $N_{\tilde\rho}$ is a constant, dependent upon $\tilde\rho$, given
in \eqref{eq:E29}.
As in the last section, it is straightforward to see that the 
$\nu_i^a$ are indeed a sum of closed cycles on the foliated
surface. To see this notice that they can be re-expressed in terms of
the closed cycles $\hat A_i^a(\tilde{\Bomega};p)=
A_i^a(\tilde{\Bomega};p)-A_i^a(\sigma_{i_p}(\tilde{\Bomega});p)$. One
can show, in a
way completely analogous to section 3, that the integrals of $\lambda$
around the $\nu_i^a$ are independent of the representation $\tilde\rho$.

As for the simply-laced groups, it is straightforward to determine the
weak coupling limit of the
integral around the cycle $\nu_i^1-\nu_i^2$. One finds that
\begin{equation}
\left(Q_i^1-Q_i^2\right)\cdot A=\Balpha_i\cdot\tilde{\Bvphi}_{\mu=0}(0),
\end{equation}
Hence, as before, 
\begin{equation}
Q_i^1-Q_i^2=(\Bzero,\Balpha_i),\label{eq:E40}
\end{equation} 
is purely electrically charged and in
the weak coupling limit $\tilde{\Bvphi}(0)$ 
is proportional to the classical Higgs VEV.

\section{The Strong Coupling Monodromies: Non-Simply-Laced}

In this section, we calculate the strong coupling monodromies for the 
non-simply-laced cases. First of all, let us write the intersection 
of the closed cycles $\hat A_i^a(\tilde{\Bomega};p)$ in the form
\begin{equation}
\hat A_i^a(\tilde{\Bomega};p)\circ
\hat A_j^b(\tilde{\Bomega};q')={\cal I}_{ij}^{ab}(p-q)
\left(\delta_{\tilde{\Bomega},\tilde{\Bomega}'}-\delta_{\sigma_{i_p}(\tilde{\Bomega}),
\tilde{\Bomega}'}-\delta_{\tilde{\Bomega},\sigma_{j_q}(\tilde{\Bomega}')}+\delta_{\sigma_{i_p}(
\tilde{\Bomega}),\sigma_{j_q}(\tilde{\Bomega}')}\right),
\end{equation}
where the intersection form ${\cal I}_{ij}^{ab}(p-q)$, for $a=b=1$, is
equal to ${\cal I}_{ij}(p-q)$ defined in the last section. 

We now apply the Picard-Lefshetz theorem to find the monodromies
around the vanishing cycles as in section 4. One finds
\begin{equation}
{\mathcal M}_{\nu_i^a}\left(\nu_j^b\right)=\nu_j^b+
{1\over2N_{\tilde\rho}}\sum_{\tilde{\Bomega}}\sum_{p,q=1}^\tau
\left(\tilde{\Balpha_{j_1}}\cdot\pi^{p-q}(\tilde{\Balpha_{i_1}})\right)
\left(\tilde{\Balpha_{i_p}}\cdot\tilde{\Bomega}\right){\cal I}_{ji}^{ba}(q-p)
\hat A_i^a(\Bomega;p).
\end{equation}
In order to simplify the above expression, we have to consider the
various cases that arise. If $i,j\in L$ then
$\tilde{\Balpha_{j_q}}\cdot\tilde{\Balpha_{i_p}}=
\tilde{\Balpha_{j_1}}\cdot\pi^{p-q}(\tilde{\Balpha_{i_1}})\neq0$ only when
$p=q$, in which case
\begin{equation}
{\mathcal M}_{\nu_i^a}\left(\nu_j^b\right)=\nu_j^b+
{\cal I}_{ji}^{ba}(0)\tilde{\Balpha}_{j_1}\cdot\tilde{\Balpha}_{i_1}
\nu_i^a.
\end{equation}
For the other three possible cases, where at least $i$ or $j$ or both
$\in S$, one finds
\begin{equation}
{\mathcal M}_{\nu_i^a}\left(\nu_j^b\right)=\nu_j^b+
\left(\sum_{p=0}^{\tau-1}{\cal I}_{ji}^{ba}(p)\right)
2\tilde{\Balpha}_{j_1}\cdot\tilde{\Balpha}_{i_1}\nu_i^a.
\end{equation}
We can further simplify the  results above and writing them in terms of
the intersection form ${\cal I}_{ij}(p)$ and in terms of the inner
product of the simple roots of $g$:
\begin{equation} 
{\mathcal M}_{\nu_i^a}\left(\nu_j^b\right)=
\nu_j^b+\left(\xi_{ba}(\Balpha_i^2/\Balpha_j^2)+D_{ji}\right)
{2\Balpha_i\cdot\Balpha_j\over\Balpha_i^2}
\nu_i^a.
\end{equation}
In the above, $D_{ji}$ is defined as follows:
\begin{align}
&j\in L:\qquad D_{ji}=\begin{cases}{\cal I}_{ji}(0)&i\in L\\
{1\over\tau}\sum_{p=0}^{\tau-1}{\cal I}_{ji}(p)&i\in S\\
\end{cases}\notag \\ 
&j\in S:\qquad D_{ji}=\sum_{p=0}^{\tau-1}{\cal I}_{ji}(p).\\
\end{align}
The matrix $\xi_{ba}(x)$ is defined as
\begin{equation}
\xi_{ba}(x)=\begin{pmatrix} 0&x\\ -1&x-1\\ \end{pmatrix}
\end{equation}

The charges must therefore have symplectic inner products
\begin{equation}
Q_j^b\cdot
Q_i^a=\left(\xi_{ba}(\Balpha_i^2/\Balpha_j^2)+D_{ji}\right)
{2\Balpha_i\cdot\Balpha_j\over\Balpha_i^2},\label{eq:E42}
\end{equation}
and be further constrained by \eqref{eq:E40}. Notice that the
expression on the right-hand-side of \eqref{eq:E42} is antisymmetric
under the interchange of $i$ and $j$, and $a$ and $b$, as it should be
for consistency. The charges are determined in the form
\begin{equation}
Q_i^1=\left(\Balpha_i^\vee,p_i\Balpha_i\right),\qquad
Q_i^2=\left(\Balpha_i^\vee,(p_i+1)\Balpha_i\right),\label{eq:E43}
\end{equation}
where, for $i$ and $j$ such that $\Balpha_i\cdot\Balpha_j\neq0$,
\begin{equation}
{\Balpha_i^2\over\Balpha_j^2}p_i-p_j=D_{ji}.\label{eq:E41}
\end{equation}
Notice immediately that the magnetic charges are indeed co-roots of
$g$ as required by the DSZ quantization condition.
The equations \eqref{eq:E41} always admit a solution up to an overall integer
ambiguity 
\begin{equation}
p_i\rightarrow p_i+\begin{cases}n&i\in L\\ n\tau&i\in S.\\ \end{cases}
\end{equation}
This reflects the usual ambiguity in the number of times that the
contours as a whole wind around the origin. 
For the $G_2$ example of Fig.~4, the charges are
\begin{equation}
Q_1^1=\left(\Balpha_1^\vee,n\Balpha_1\right),\qquad
Q_2^1=\left(\Balpha_i^\vee,(3n-1)\Balpha_2\right),
\end{equation}
for integer $n$.

If we re-define $A_i^1(p)$, with $i\in L$, to have an
additional winding around the origin this corresponds to a shift
$p_i\rightarrow p_i+1$. Similarly, if we redefine $A_i^1(p)$, $i\in S$, 
so that it connects $z_i^-(p)$
to $z_i^+(p+1)$, rather than $z_i^+(p)$, then this also corresponds to a
shift $p_i\rightarrow p_i+1$. Hence, a dyon with charge
$(\Balpha_i^\vee,n\Balpha_i)$ is related to a vanishing cycle which is
the lift of $A_i^1(p)+(n-p_i)C$, when $i\in L$, and  
$A_i^1(p)+\sum_{a=0}^{n-p_i-1}C(p+a\,{\rm mod}\tau)$, when $i\in S$.
As a consequence of this, there is
a vanishing cycle for each of the dyons of charge
$(\Balpha_i^\vee,n\Balpha_i)$, for any $n\in{\mathbb Z}$. As before, this
reflects the democracy of dyons.
 
\section{Discussion}

Now that we have shown that the strong coupling monodromies can be
explained by dyons, we can subject the Martinec-Warner construction to
two additional non-trivial tests. 

The first of these involves showing that the correct weak coupling
monodromies are produced. A 
monodromy at weak coupling $M_i$ corresponds to a path 
for which $z_i^\pm\rightarrow\exp(\pm2\pi i)z_i^\pm$, for the simply-laced
algebras, and $z_i^\pm(p)\rightarrow \exp(\pm2\pi i\tau_i/\tau)z_i^\pm(p)$,
for the non-simply-laced algebras. These transformations
can be achieved without leaving the weak coupling regime. Such a path
corresponds precisely to the encircling the pair of strong coupling
singularities corresponding to the
vanishing cycles $\nu_i^a$, $a=1,2$. The picture is then a rather simple
generalization of the SU(2) case discussed by Seiberg and Witten \cite{SW}.
Being careful with the order, one finds that
\begin{equation}
M_i=M_{Q_i^2}M_{Q_i^1}.\label{eq:E45}
\end{equation}
It is a simple matter to verify that the right-hand-side is equal to the
weak coupling monodromy in \eqref{eq:E10}, when computed
using \eqref{eq:E11}, and the facts that
$Q_i^1=(\Balpha_i^\vee,n\Balpha_i)$, for some $n\in{\mathbb Z}$, and
$Q_i^1-Q_i^2=(\Bzero,\Balpha_i)$.

The second non-trivial test is to compare the spectrum of dyons which
are responsible for the strong coupling singularities with those
present at weak coupling. There is no absolute guarantee that dyons
present at weak coupling will survive without decay all the way to the
strong coupling singularities, although it can be shown that they do
in SU(2) \cite{SW}. 
Nevertheless, the match is perfect because the weak coupling
coupling spectrum consists of dyons of charge
$(\Balpha_i^\vee,n\Balpha_i)$, for $n\in{\mathbb Z}$, precisely the
ones responsible for the strong coupling singularities. The weak
coupling spectrum also consists of dyons whose magnetic charges are
non-simple co-roots. The spectrum  of such dyons is complicated by the
fact the allowed electric charges vary in different cells of moduli
space separated by surfaces on which these dyons decay \cite{TJH3,TJH4}. 
However, such
dyons are related by products of weak coupling monodromies to those 
dyons whose magnetic charge is a simple root. Hence
there is a true democracy, since for the dyons whose
magnetic charge is a non-simply root, one can first
follow a path which undoes the semi-classical monodromy and then
proceed to a singularity.

\vspace{1cm}
I would like to thank Nick Warner for useful discussions.
I would also like to thank PPARC for an Advanced Fellowship.

\end{document}